\documentclass[twocolumn,showpacs,floatfix,prb]{revtex4}

\usepackage{amssymb}
\usepackage{graphicx}
\usepackage{dcolumn}
\usepackage{bm}

    \font\tenbifull=cmmib10 
    \font\tenbimed=cmmib7
    \font\tenbismall=cmmib5
\def\bmit{\fam9 }
\mathchardef\bbGamma="7000 \mathchardef\bbDelta="7001
\mathchardef\bbPhi="7002 \mathchardef\bbAlpha="7003
\mathchardef\bbXi="7004 \mathchardef\bbPi="7005
\mathchardef\bbSigma="7006 \mathchardef\bbUpsilon="7007
\mathchardef\bbTheta="7008 \mathchardef\bbPsi="7009
\mathchardef\bbOmega="700A \mathchardef\bbalpha="710B
\mathchardef\bbbeta="710C \mathchardef\bbgamma="710D
\mathchardef\bbdelta="710E \mathchardef\bbepsilon="710F
\mathchardef\bbzeta="7110 \mathchardef\bbeta="7111
\mathchardef\bbtheta="7112 \mathchardef\bbiota="7113
\mathchardef\bbkappa="7114 \mathchardef\bblambda="7115
\mathchardef\bbmu="7116 \mathchardef\bbnu="7117
\mathchardef\bbxi="7118 \mathchardef\bbpi="7119
\mathchardef\bbrho="711A \mathchardef\bbsigma="711B
\mathchardef\bbtau="711C \mathchardef\bbupsilon="711D
\mathchardef\bbphi="711E \mathchardef\bbchi="711F
\mathchardef\bbpsi="7120 \mathchardef\bbomega="7121
\mathchardef\bbvarepsilon="7122 \mathchardef\bbvartheta="7123
\mathchardef\bbvarpi="7124 \mathchardef\bbvarrho="7125
\mathchardef\bbvarsigma="7126 \mathchardef\bbvarphi="7127

\def\boldsigma{\bmit\bbsigma}

\begin{document}
\title{Strongly correlated regimes in a double quantum-dot device}  
\author{P. S. Cornaglia}
\author{D. R. Grempel}
\affiliation{CEA-Saclay, DSM/DRECAM/SPCSI, B\^at. 462, F-91191 
Gif-sur-Yvette, France}
\begin{abstract}
The transport properties of a double quantum-dot device
 with one of the dots coupled to perfect conductors
 are analyzed using the numerical renormalization group technique and
 slave-boson mean-field theory. The coupling between the dots strongly
 influences the transport through the system leading to a
 non-monotonic dependence of the conductance as a function of the temperature
 and the magnetic field. For small inter-dot coupling and parameters such that
 both dots are in
 the Kondo regime, there is a two-stage screening of the dot's magnetic
 moments that is reflected in the conductance. 
In an intermediate temperature regime 
Kondo correlations develop on one of the dots
 and the conductance is enhanced. At low temperatures the Kondo effect
 takes place on the second dot leading to a singlet ground state
in which the conductance
is  strongly suppressed. 
\end{abstract}
\maketitle 

Since the prediction~\cite{Glazman1988,Ng1988} of the occurrence of the 
Kondo effect in a
single quantum dot (QD) device and its subsequent
experimental observation\cite{Gordon1998}, several single and double quantum
dot devices (DQDD) have been studied both
theoretically~\cite{Georges1999,Busser2000} and experimentally.~\cite{Chen2004,Wiel2003}
The interest in this systems stems form their potential 
applications to quantum and classical computing
\cite{Loss1998,DiCarlo2004} and their usefulness as model systems 
 for the study of the physics of strongly correlated electrons.\cite{Chen2004}

In its simplest form, the Kondo effect consists in the screening of a
 localized spin
$1/2$ magnetic moment antiferromagnetically coupled to a conduction-electron
 band  and appeared 
originally in the context of magnetic impurities in a
metallic host.  In a QD device Coulomb repulsion and spatial confinement
 result in charge quantization  
and its transport properties are dominated by the Coulomb blockade. 
When the charge on the QD is close to an odd integer 
the Kondo efect takes place and it results in perfect transmission 
through the system at sufficiently low temperatures 
in a wide range of gate voltages.

When two dots are coupled, a richer and more complicated behavior may arise 
 \cite{Georges1999,Busser2004,Chen2004} including the possibility 
of observing quantum phase transitions. \cite{Vojta2002}

\begin{figure}[tbp]
\includegraphics[width=5cm,clip=true]{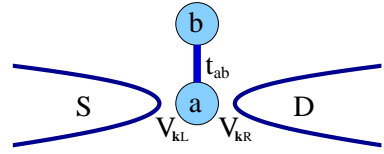}
\caption{Schematic representation of the device. A QD molecule
with  only one of the dots coupled to the leads.}
\label{fig:device}
\end{figure}

In this paper we present a detailed study of a DQDD in which only one
of the dots is coupled to the leads (see Figure \ref{fig:device}).
We discuss the case in which the QDs are very small and have a  
single relevant energy level at energies $\varepsilon_a$ and 
 $\varepsilon_b$ and large
charging energies $U_a$ and $U_b$, where the indices $a$ and $b$ denote the two dots. The position of  energy levels of the dots and 
the strength of the coupling between them $t_{ab}$
can be tuned by applying gate voltages in semiconductor devices.
\cite{DiCarlo2004}

The Hamiltonian of the DQDD is 
\begin{equation} \label{eq:hamilt}
  H=H_D+H_E+H_{D-E}\;,
\end{equation}
where
\begin{eqnarray}
H_D &=& \sum_{\ell=a,b} \left[U_\ell n_{\ell \uparrow} n_{\ell
\downarrow}+ \varepsilon_\ell (n_{\ell \uparrow} +n_{\ell
\downarrow})\right]- \nonumber\\ && t_{ab}\sum_\sigma
\left(d^\dagger_{a\sigma} d_{b\sigma} +d^\dagger_{b\sigma}
d_{a\sigma}\right).
\end{eqnarray}
describes the $a-b$ molecule and 
\begin{equation}
 H_{E}=\sum_{{\bf k},\sigma,\alpha} \varepsilon_{\alpha}({\bf k})\;
c^{\dagger}_{{\bf k} \sigma \alpha} c^{}_{{\bf k} \sigma
\alpha}\;\left(\alpha={\rm L,R}\right)\;,\nonumber
\end{equation}
is the Hamiltonian of two non-interacting source and drain leads.
The coupling between the molecule and the leads is described by the last
 term in the Hamiltonian, 
\begin{equation}
H_{D-E}=\sum_{{\bf k},\sigma,\alpha}\;V_{{\bf
k}\alpha}\left(d^{\dagger}_{a\sigma}
\;c^{}_{{\bf k}\sigma\alpha} + c^{\dagger}_{{\bf
k}\sigma\alpha}\;d^{}_{a\sigma} \right)\;. \nonumber
\end{equation}

For symmetric leads, the conductance through the system is~\cite{Meir1992}
\begin{equation}\label{eq:cond}
G(T)=\frac{e^{2}}{h} \Delta \pi
\sum_{\sigma}\int_{-\infty}^{\infty} d\omega \left(-\frac{\partial
f(\omega )}{\partial \omega }\right)\rho^{\sigma}_{aa}(\omega ),
\end{equation}
where $\rho^{\sigma}_{aa}(\omega)$ is the local electronic density
of states on dot $a$.  Here, $\Delta=2\pi \rho_0\langle V_{\bf
k}^2\rangle$ where $\rho_0$ is the electronic density of states of
the electrodes at the Fermi level, the brackets denote the average
over the Fermi surface, and $f(\omega)$ is the Fermi function.

We solved the above model using the numerical renormalization group
 technique as well as an approximate analytical method
 that reproduces the main features of the numerical solution.
 Some exact results, that follow from the application of
 Luttinger's theorem to this system,  
 are also given.

Our main findings are the following. The coupling $t_{ab}$ is a
 relevant perturbation and the low-energy physics of the DQDD is
 fundamentally 
 different from that of a single QD device. 
 The zero-temperature conductance of the system vanishes
 when the total charge of the device is an even integer and it is perfect
 when the charge is an odd integer. This is an exact result. 

 When both dots are nearly half-filled there are 
two different regimes for the temperature-dependence of the
 conductance depending on whether  $t_{ab}$ is small
 or large in a sense that will be made precise below. 
 For small $t_{ab}$,  
 a two-stage analog of the Kondo effect 
 takes place with decreasing temperature. In the first stage,
 the magnetic moment of dot $a$ is quenched by the electrons
 of the leads below some temperature. 
In the second stage, that occurs at a much lower temperature, 
 the magnetic moment of dot $b$ is in turn quenched by electrons
 that lie within the narrow Kondo resonance around the Fermi level.
 The two-stage Kondo effect results in a non-monotonic behavior
 of the conductance as a function of both gate voltage and magnetic field. 
In the regime of large $t_{ab}$, the magnetic moments of the dots form a
 tightly bound singlet weakly coupled to the leads and the conductance
varies monotonically at low temperature.

When the charge on the device is varied by application of a
 gate voltage $V_g$, the zero-temperature
 conductance remains small in an extended interval of values of  $V_g$. 
Conversely, in general, nearly  perfect  
 conductance is only observed in a very restricted
 range of values of the gate voltage.  

 A surprisingly rich variety of possibilities exists
 for the dependence of  the conductance upon
 gate voltage and temperature depending upon
 the various parameters of the model.    

The rest of the paper is organized as follows.
In Section \ref{sec:Fermi} we derive exact results for the dependence
 of the zero-temperature conductance of DQDD upon its charge and
 magnetization. The solution of the model within
 slave-boson mean-field theory is presented in  
Section \ref{sec:SBMFT}. Section \ref{sec:NRG} contains the description of our numerical results.   Finally, we summarize our conclusions in Section 
\ref{sec:conc}.

\section{Exact results}\label{sec:Fermi}

In the configuration depicted in Fig.~\ref{fig:device}
the ground state of a DQDD is expected to be a Fermi liquid. In
this case, many exact results for the zero-temperature
conductance of the device can be obtained using a simple generalization of
Luttinger's theorem that we derive in the following.

Since electron-electron interactions do not involve the electronic
states of the  leads, the latter may be integrated out from the
start. All the properties of the DQDD can then be expressed in
terms of the reduced $2\times 2$ retarded Green-function matrix
\begin{equation}
{\bf G}^{-1}(\omega)=
{\bf G}_0^{-1}(\omega) - {\bf \Sigma}(\omega)\;,
\label{green}
\end{equation}
where ${\bf G}_0(\omega)$ is the non-interacting Green function,
\begin{eqnarray}
{\bf G}_0^{-1}(\omega)=\left(
\begin{array}{cc} \omega - \epsilon_a + i \Delta & t_{ab} \\
t_{ab} & \omega - \epsilon_b
\end{array}
\right)\;,
\label{green0}
\end{eqnarray}
and ${\bf \Sigma}(\omega)$ is the self-energy matrix.

The total number of electrons with spin $\sigma$ in the dots at
$T=0$ is given by
\begin{equation}
n^\sigma\equiv n_a^\sigma + n_b^\sigma =-{1 \over \pi} \int_{-
\infty}^0\;d\omega\; \textrm{Im}\left[\textrm{Tr}{\bf
G}(\omega)\right]\;, \label{occupations}
\end{equation}
an expression that can be rewritten in the form
\begin{eqnarray}
n_{\sigma} &= -{1 \over \pi}\;\textrm{Im}\;
\int_{-\infty}^0\;d\omega\; {\partial \over \partial
\omega}\textrm{Tr}\ln {\bf G}^{-1}(\omega) \nonumber\\
 &{}- {1 \over \pi}\;\textrm{Im} \int_{-\infty}^0\;d\omega\;
 \textrm{Tr} \left[ {\bf G}(\omega) {\partial \over \partial
\omega}{\bf \Sigma}(\omega)\right]\;,
\label{intermediate}
\end{eqnarray}
using the equality
\begin{equation}
\textrm{Tr} {\bf G}(\omega) = {\partial \over \partial
\omega}\;\textrm{Tr}\ln {\bf G}^{-1}(\omega) +  \textrm{Tr}\left[
{\bf G}(\omega) {\partial \over \partial  \omega}{\bf
\Sigma}(\omega)\right]\;, \label{equivalence}
\end{equation}
that can be easily checked using Eqs.~(\ref{green}) and
(\ref{green0}).

The second integral on the right-hand side of
Eq.~(\ref{intermediate}) vanishes order by order in perturbation
theory in $U_{a,b}$~\cite{Abrikosov} which leads to
\begin{eqnarray}
n_{\sigma}= {1\over \pi}
\left[\varphi(-\infty)-\varphi(0)\right]\;,
\end{eqnarray}
where
\begin{eqnarray}
\varphi(\omega) &=& \textrm{Im}\ln \Big\{\left[\omega - \epsilon_a -
\Sigma_{aa}(\omega) + i \Delta\right]\nonumber\\
&\times& \left[\omega - \epsilon_b -\Sigma_{bb}(\omega)\right] -
\left[t_{ab}-\Sigma_{ab}(\omega)\right]^2\Big\}\;.
\label{fase}
\end{eqnarray}
Introducing the renormalized parameters ${\tilde{\epsilon}}_{a}=
\epsilon_a + \Sigma_{aa}(0)$, ${\tilde{\epsilon}}_{b}= \epsilon_b
+ \Sigma_{bb}(0)$, and ${\tilde{t}}_{ab}= t_{ab} - \Sigma_{ab}(0)$
we find:
\begin{eqnarray}
n_{\sigma}=2 - {1\over \pi} \textrm{Im}\ln \left[
\tilde{\epsilon}_a \tilde{\epsilon}_b  - {\tilde{t}}_{ab}^2 - i
\Delta \tilde{\epsilon}_b\right]\;.
\label{occ-ren}
\end{eqnarray}
For a Fermi-liquid ground state the self-energy matrix at the
Fermi energy is purely real and so are the renormalized
parameters. A straightforward calculation then yields:
\begin{eqnarray}
n_\sigma = \frac{1}{\pi}\left\{
\begin{array}{ll}
\delta\;, & \tilde{\epsilon}_a\;\textrm{and}\;\tilde{\epsilon}_b
> 0\;,\;\;\tilde{\epsilon}_a\;\tilde{\epsilon}_b > \tilde{t}_{ab}^2
\;\\ 2\pi +\delta\;, &
\tilde{\epsilon}_a\;\textrm{and}\;\tilde{\epsilon}_b < 0\;,
\;\;\tilde{\epsilon}_a\;\tilde{\epsilon}_b > \tilde{t}_{ab}^2 \;\\
\pi + \delta\;, & \textrm{otherwise}
\end{array}
\right. \;, \label{evaluation-phase}
\end{eqnarray}
where \begin{equation}
\delta =  \arctan\left( {\tilde{\epsilon}_b
\Delta \over \tilde{\epsilon}_a
 \tilde{\epsilon}_b - \tilde{t}_{ab}^2 }
\right)\;.
\label{delta}
\end{equation}
In all cases we have the relationship
\begin{eqnarray}
\left({\tilde{\epsilon}_b \Delta \over \tilde{\epsilon}_a
 \tilde{\epsilon}_b - \tilde{t}_{ab}^2 }\right)^2 = \tan^2 \left(\pi
n_{\sigma} \right)\;. \label{tangent2}
\end{eqnarray}
Expressions~(\ref{evaluation-phase})-(\ref{tangent2}) are
identical to those that one would obtain for the non-interacting
problem defined by Eq.~(\ref{green0}) with the renormalized real
parameters replacing the bare ones.

Similarly, the $a$-dot spectral density for spin $\sigma$ at the
Fermi energy is given by
\begin{eqnarray}
\rho^\sigma_{aa}(0) = {1\over \pi} {\tilde{\epsilon}_b^2 \Delta
\over \left(\tilde{\epsilon}_a \tilde{\epsilon}_b -
\tilde{t}_{ab}^2 \right)^2 + \left(\tilde{\epsilon}_b
\Delta\right)^2 }\;, \label{rho}\end{eqnarray} which, using
Eq.~(\ref{tangent2}), can be cast in the form
\begin{eqnarray}
\rho^\sigma_{aa}(0)= {1\over \pi \Delta}  \sin^2\left(\pi n_{\sigma}
\right)\;.
\label{rhof}
\end{eqnarray}
Equations~(\ref{eq:cond}) and (\ref{rhof}) lead to the following
expression for the conductance:
\begin{eqnarray}\label{eq:condFL}
g\equiv{G \over G_0} =\frac{1}{2} \sum_{\sigma}\;\sin^2\left[\pi
\left(n_{a}^\sigma + n_{b}^\sigma \right)\right]\;,
\end{eqnarray}
where $G_0=2 e^2/h$ is the quantum of conductance.

In the absence of a magnetic field
$n_\ell^\sigma=n_\ell^{-\sigma}$ and $g=\sin^2\left[{\pi \over 2}
\left(n_{a} + n_{b} \right)\right]$ where $n_{\ell} = \sum_\sigma
n_{\ell}^\sigma$. The zero-temperature conductance thus vanishes
when the total number of electrons in the DQDD is even. This
occurs in particular when $\varepsilon_\ell = -U_\ell/2$ for both
dots. At this point the Hamiltonian is electron-hole symmetric and
$n_a+n_b=2$. Applying a gate voltage $V_g$ that shifts the levels
of the dots, the charge of the device varies continuously between
$n_a+n_b=0$ (for $V_g$ large and negative) and $n_a+n_b=4$ (for
$V_g$ large and positive). The conductance $g(V_g)$ vanishes at
$V_g \to \pm \infty$ and $V_g=0$ and has two maxima where it
reaches the value $g=1$ at the values of the gate voltage for
which $n_a+n_b=1$ or 3.

In the presence of a magnetic field $B$ the dots polarize and a
magnetization $m=\frac{1}{2} (n_{\uparrow}-n_{\downarrow})$
appears. In the regime $n_a + n_b=2$ Eq.~(\ref{eq:condFL}) becomes
\begin{equation}
 g = \sin^2(\pi m).
\label{eq:luttinger-m}
\end{equation}
In the low field limit $g$ vanishes quadratically with $B$,
\begin{equation}
 g \approx \pi^2 m^2 \approx \pi^2 B^2 \chi^2 = (B/B^\star)^2
\label{g-de-B}
\end{equation}
were $\chi$ is the spin susceptibility and $B^\star = 1/(\pi
\chi)$ defines a characteristic magnetic field. For very large $B$
the dots become fully polarized, $m \to 1$, and the conductance
also vanishes. At some intermediate field for which $m = 1/2$ the
conductance has a maximum at which $g=1$.

\section{Slave Boson mean field theory}\label{sec:SBMFT}
In this Section we present a mean-field slave-boson
treatment~\cite{largeN,Coleman87,Hewson} (SBMFT) of the problem.
For simplicity we restrict ourselves to the electron-hole
symmetric case in the most interesting regime in which the Coulomb
repulsion on the dots $U_{a,b} \gg \Delta$. In this case the
occupation of each dot
 is $n_{\ell} \sim 1$ and the low-energy
excitations of the DQDD are spin fluctuations. These are described
by the effective Hamiltonian
\begin{equation}\label{eq:Hkondo}
H_K = J_{ab} {\bf S}_a\cdot {\bf S}_b +  J {\bf S}_a\cdot {\bf
s}_0  + H_E.
\end{equation}
Here, ${\bf S}_{\ell}$ with $\ell = a,b$ are spin operators
associated to the dots, and ${\bf s}_{0}=\frac{1}{2}
\sum_{s,s^\prime} c^{\dagger}_{0
s}{\boldsigma}_{s,s^\prime}c^{}_{0 s^\prime}$ is the electron spin
density on the orbital coupled to dot $a$. To leading order the
coupling constant are $J_{ab}=8 t_{ab}^2/(U_a+U_b)$ and $J = 8
\Delta/\pi U_a$.

Following standard methods~\cite{largeN,Coleman87,Hewson} we represent the
 spin operators as ${\bf S}_{\ell}=\frac{1}{2}
\sum_{s,s^\prime} f^{\dagger}_{\ell
s}{\boldsigma}_{s,s^\prime}f^{}_{\ell s^\prime}$ in terms of two
pseudo-fermions constrained by the conditions
$\sum_{s}f^{\dagger}_{\ell s}f^{}_{\ell s}=1$. The biquadratic
interactions between the pseudo-fermions generated by the
spin-spin interactions are decoupled introducing two Bose fields,
$B_a$ (conjugate to the amplitude $\sum_s f^{\dagger}_{a s}
c^{}_{0 s}$ and $B_{ab}$ (conjugate to the amplitude $\sum_s
f^{\dagger}_{a s} f^{}_{b s}$) and the constraints on occupations
are implemented through Lagrange multipliers fields
$\lambda_{\ell}$. The free energy expressed in terms of the Bose
fields has a saddle point at which the latter condense, $\langle
B_{a}\rangle = \langle f^{\dagger}_{a \sigma} c^{}_{0
\sigma}\rangle$, $\langle B_{a b}\rangle = \langle f^{\dagger}_{b
\sigma} f^{}_{a \sigma}\rangle$, and the Lagrange multipliers
vanish. The solution obtained by retaining only the saddle-point
contribution to the free-energy is exact if the spin symmetry is
extended from SU(2) to SU($N$) and the limit $N\to \infty$ is
taken. In the physical $N=2$  case the SBMFT approach provides a
simple and yet qualitatively accurate description of the
low-energy properties of the system.

At the saddle point the
effective Hamiltonian is
\begin{equation}
H_\textrm{eff} =\widetilde{J}_{ab}
\sum_{\sigma}(f^{\dagger}_{a\sigma} f^{}_{b\sigma} + h.c. ) +
\widetilde{J} \sum_{\sigma}(f^{\dagger}_{a\sigma} c^{}_{0\sigma} +
h.c. ) +H_E,
\end{equation}
where the renormalized couplings are determined self-consistently
by the equations \vspace{0.2cm}
\begin{eqnarray}
\label{eq:Jabtilde} \frac{\pi \widetilde{J}_{ab}}{J_{ab}} &=&
\int_{-D}^D
d\omega\;f(\omega)\;\textrm{Im}\;{\cal G}_{ab}(\omega)\;,\\
\label{eq:Jtilde}
 \frac{1}{J\rho_0} &=& - \;\int_{-D}^D
d\omega\;f(\omega)\;\textrm{Re}\;{\cal G}_{aa}(\omega)\;,
\end{eqnarray}
where
\begin{eqnarray}
{\cal G}^{-1}(\omega)=\left(
\begin{array}{cc} \omega + i \widetilde{\Delta} & - \widetilde{J}_{ab} \\
- \widetilde{J}_{ab} & \omega
\end{array}
\right)\;, \label{greenSBMFT}
\end{eqnarray}
and $D$ is the half-width of the conduction band.

The system of equations (\ref{eq:Jabtilde})-(\ref{greenSBMFT}) can
be easily solved numerically. However, we shall focus below on
some relevant limiting cases for which analytical expressions  can
be obtained.

\subsection{Decoupled dots}
For $J_{ab}=\widetilde{J}_{ab}=0$ the model
reduces to the well known Kondo Hamiltonian. The results for this
model in the SBMFT approximation are well known. \cite{Hewson} In
the weak-coupling limit $J\rho_0 \ll 1 $ the solution of
Eq.~(\ref{eq:Jtilde}) at $T=0$ is $\widetilde{\Delta}_0 =
De^{-1/\rho_0J}$. At finite temperatures solutions with non-zero
$\widetilde{\Delta}_0$ only exist below a temperature $T_c = c_1 D
e^{-1/\rho_0J}$, where $c_1$ is a constant of order $1$. This
approximation thus gives a spurious transition at
which the spins and the conduction electrons decouple. In the
exact solution of the Kondo model this sharp transition is
replaced by a crossover at the Kondo temperature $T_K^0 \propto D
e^{-1/\rho_0J}$. We thus identify $T_c$ with $T_K^0$.

The quasiparticle density of states is
\begin{equation}\label{eq:rho0}
\rho_{aa}(\omega) = {1\over \pi}
\frac{\widetilde{\Delta}_0}{\omega^2 + \widetilde{\Delta}_0^2 }\;,
\end{equation}
describing a resonance of width $\widetilde{\Delta}_0$ at the
Fermi level. The presence of this resonance leads to perfect
conductance, $g=1$, at zero temperature.
The dot's magnetic susceptibility is $\chi_{aa}\propto
1/\left(\pi \widetilde{\Delta}_0\right)$.

\subsection{Weakly coupled dots}
In the limit in which the coupling between the dots is weak,
$J_{ab}\ll \widetilde{\Delta}_0$ , the solutions of the equations
at $T=0$ to leading order in
$\widetilde{J}_{ab}/\widetilde{\Delta}$ are
\begin{eqnarray}
\widetilde{J}_{ab}&\sim& \widetilde{\Delta} \exp\left( - \frac{\pi
\widetilde{\Delta}}{2 J_{ab}}\right)
\\ \nonumber \\
\widetilde{\Delta}&\sim& \widetilde{\Delta}_0 \left[1 - 2
\left(\frac{\widetilde{J}_{ab}}{\widetilde{\Delta}_0}\right)^2
 \ln \left(\frac{\widetilde{\Delta}_0}{\widetilde{J}_{ab}}
\right) \right].
\label{Delta-wc}
\end{eqnarray}
While $\widetilde{\Delta} \approx\widetilde{\Delta}_0$
remains essentially unchanged in this
regime, the effective coupling between the dots is strongly
suppressed. With increasing temperature $\widetilde{J}_{ab}$
further decreases and vanishes at a temperature
\begin{equation}\label{eq:T0}
T_0 = c_2 \widetilde{\Delta}_0 \exp(-\frac{\pi
  \widetilde{\Delta}_0}{J_{ab}}) \propto
\frac{\widetilde{J}_{ab}^2}{\widetilde{\Delta}_0 }\;,
\end{equation}
where $c_{2} \sim 1$ depends weakly on the parameters. As above, this
transition temperature should be interpreted as a crossover
temperature below which the magnetic moment of the dot $b$ is
screened.

Note that $T_0 \ll T_K^0$ which suggests a two-stage screening
process. \cite{Vojta2002}  First, the magnetic moment of dot $a$ is screened
by the electrons of the leads below $T_K^0$. Then, at the much
lower temperature $T_0$, the magnetic moment of dot $b$ is
screened by the heavy quasiparticles of the local Fermi liquid that forms
 on dot $a$ for $T\ll T^0_K$. 
The form of $T_0$ supports
this physical picture as it corresponds precisely  to the
expression of the Kondo temperature of a magnetic moment screened
by electrons of a band of width $\sim \widetilde{\Delta}_0$ 
and density of states $1/(\pi \widetilde{\Delta}_0)$. 

The quasiparticle densities of states of the dots are
\begin{eqnarray}
\rho_{aa}(\omega) &=& {1\over \pi} {\omega^2 \widetilde{\Delta} \over
\left(\omega^2 - \widetilde{J}_{ab}^2 \right)^2 + \left(\omega
\widetilde{\Delta}\right)^2 }\;,
\label{rho_MF_aa}
\\ \nonumber\\
\rho_{bb}(\omega) &=& {1\over \pi} {\widetilde{J}_{ab}\widetilde{\Delta} \over
\left(\omega^2 - \widetilde{J}_{ab}^2 \right)^2 + \left(\omega
\widetilde{\Delta}\right)^2 }\;,
\label{rho_MF_bb}\end{eqnarray}
respectively.

The first of these equations determines the conductance through the device.
In the temperature range $T_0 \ll T \ll T^0_K$, where $\widetilde{J}_{ab}$
 is irrelevant, we recover the usual Kondo resonance
\begin{eqnarray}
\rho_{aa}(\omega) \sim {1\over \pi}
\frac{\widetilde{\Delta}}{\omega^2 + \widetilde{\Delta}^2 }\;.
\end{eqnarray}
This leads to a large conductance $g \approx 1$.

For $T \ll T_0$ and $\omega \ll \widetilde{J}_{ab}$,
\begin{eqnarray}
\rho_{aa}(\omega) \sim \frac{1}{\pi \widetilde{\Delta}_0}
\;\frac{\omega^2}{ \omega^2 + \left(\widetilde{J}_{ab}^2/\widetilde{\Delta}_0
\right)^2
}\;.
\end{eqnarray}
Therefore, $\rho_{aa}$ vanishes at the Fermi energy where it has a
dip of width $\propto T_0$. This dip is the analog of the Kondo
hole that appears in the conduction-electron density of states in
the usual Kondo problem.
An immediate
consequence of the presence of this hole is that the
zero-temperature conductance of the DQDD vanishes at $T=0$.

In the same limit, the magnetic susceptibility of the DQDD is dominated by
 the contribution of dot $b$. We find
\begin{equation}
\label{chiSBMFT}
\chi \approx \chi_{bb} = \frac{\widetilde{J}_{ab}^2}{\pi 
\widetilde{\Delta}_0}
 \propto \frac{1}{T_0}\;.
\end{equation}
Therefore, the magnetic scale $B^\star$
defined in Eq.~(\ref{g-de-B}) is also determined by $T_0$.

\subsection{Strongly coupled dots}
For $J_{ab} >\widetilde{\Delta}_0$ there are no solutions of
Eqs.~(\ref{eq:Jabtilde}) and (\ref{eq:Jtilde}) with non-vanishing
$\widetilde{\Delta}$. The dots remain decoupled from the leads
and
$\widetilde{J}_{ab} = J_{ab}/2$. The ground state is a singlet and there
 is a energy gap   $ E_g = 2\widetilde{J}_{ab} = J_{ab}$
 in the electronic spectrum. 
Note that the singlet-triplet gap that results from the SBMFT treatment
 is the exact result for a pair of isolated dots.

The decoupling of the dots from the leads is
an artifact of SBMFT. Presumably, the fluctuation
 corrections neglected in this approach generate a weak coupling
 between the spin singlet and the leads but we have not checked
 this point explicitly.

\section{Numerical Results}\label{sec:NRG}

In this section we present the numerical solution of the DQDD model
 obtained using the numerical renormalization group method
(NRG)~\cite{Wilson1975,Krishnamurthy1980,Costi1994} modified to
improve the accuracy in the computation of the spectral density.
\cite{Hofstetter2000} We choose the half-bandwidth of the
conduction band $D$ as the unit of energy and consider for
simplicity the case of identical dots,
$\varepsilon_a=\varepsilon_b \equiv \varepsilon$ and $U_a=U_b
\equiv U$.

We shall first discuss the electron-hole symmetric case in order
 to make contact with the SBMFT results of the previous Section.

\subsection{Kondo screening}
\label{screening}

Figure \ref{fig:fig1} shows the NRG results for the temperature
dependence of square of the total magnetic moment of the
 DQDD  in the electron-hole symmetric case for several values of $t_{ab}$.
 The other parameters are $U=0.5$, $\varepsilon=-0.25$, and $\Delta=0.035$.

The magnetic moment $\mu$ is defined through the relationship
$\mu^2= T \chi$ where $\chi$ is the contribution of the quantum
dots to the total magnetic susceptibility of the system.~\cite{Wilson1975} 
For a
free $S$=1/2 spin, $\mu^2=1/4$.

The uppermost curve in Fig.~\ref{fig:fig1} represents the case of
decoupled dots, $t_{ab}=0$. At high temperatures (but low compared
to $U$)
 the spins of both dots are free. Then $\mu^2 \to 1/2$.
At some temperature scale there is a drop in $\mu^2$ due to the
Kondo screening of dot $a$. We identify this temperature with the
Kondo temperature $T^0_K$ (a more precise determination of $T^0_K$
will be given in the following Section). Since the spin on dot $b$
remains free down to $T=0$, $\mu^2 =1/4$ for $T \ll T_K^0$.

For any finite $t_{ab}$  the total magnetic moment in the
 ground state vanishes and $\mu^2 \to 0$ as $T\to 0$.
All the curves but the lowermost correspond to values of $t_{ab}$ such that
$J_{ab} \le T^0_K$. It can be seen that, within this range of parameters, 
 screening takes place in two stages as predicted by  SBMFT.
With decreasing temperature
 there is a first drop in $\mu^2$ at the scale $T^0_K$ followed by a plateau. A
  second drop occurs at a much lower scale that we shall identify below with
  $T_0$. The temperature at which the first drop
occurs is rather
 insensitive to the value of $J_{ab}$ in agreement
with Eq.~(\ref{Delta-wc}). Note that
  the magnetic moment varies logarithmically with temperature
 at the two steps which confirms the two stages of screening
 are of the Kondo type as stated in
the previous Section.

When $t_{ab}$ increases the steps become closer, which signals a
merging of the two scales with increasing $J_{ab}$, in agreement
with Eq.~(\ref{eq:T0}).

We found that, in this low-$J_{ab}$ regime and at low temperature, 
all the
curves collapse onto a universal curve
 if
 the temperature is scaled by an appropriate factor $T_{ab}$
 that depends on $J_{ab}$.
We expect on physical grounds $T_{ab}\propto T_0$. In order to
check this hypothesis we calculated $T_{ab}$ from the
susceptibility data using the criterion $\mu^2(T_{ab}) = 1/8$. The
results are represented with solid circles
 in the inset to Fig.~\ref{fig:fig1}. It is seen that
$T_{ab}$ varies exponentially with $t_{ab}^{-2}$
 in agreement with Eq.~(\ref{eq:T0}).
The solid line is a fit to the SBMFT expression for $T_0$. An
arbitrary multiplicative constant in the definition of $T_{ab}$
was absorbed in $c_2$ and the parameter
 $c_1$ is the proportionality factor between $T^0_K$ and
 $\widetilde{\Delta}_0$. As expected, the values of $c_1$
and $c_2$ determined by the fit are ${\cal O}(1)$.

The lowermost curve shown in Fig.~\ref{fig:fig1} corresponds to
the case $J_{ab}>T^0_K$. It is seen that it is qualitative
different from the others. In this regime the spins of the dots
form a robust singlet weakly coupled to the leads. The
susceptibility and the magnetic moment decrease exponentially in
$J_{ab}/T$.

\subsection{Spectral properties and conductance}
\begin{figure}[tbp]
\includegraphics[width=8.5cm,clip=true]{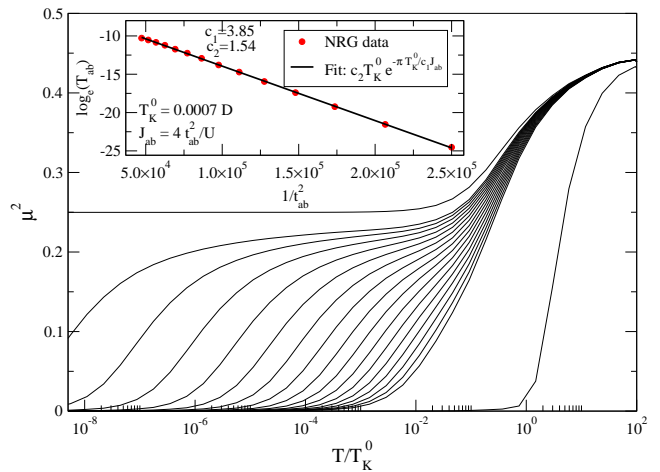}
\caption{Magnetic moment of the DQDD as a function of the
temperature for different values of the interdot coupling. From
top to bottom, $t_{ab}=0, t_{ab} = 0.002-0.0046$ in steps of 0.0002,
and $t_{ab} = 0.02$. Parameters are $U=0.5$, $\varepsilon=-0.25$ and
$\Delta=0.035$. Inset: binding energy of the $a-b$ singlet
 $T_{ab}$ estimated using the NRG (see text). The solid line is a fit
based on Eq.~(\ref{eq:T0}).} \label{fig:fig1}
\end{figure}

\begin{figure}[tbp]
\includegraphics[width=8.5cm,clip=true]{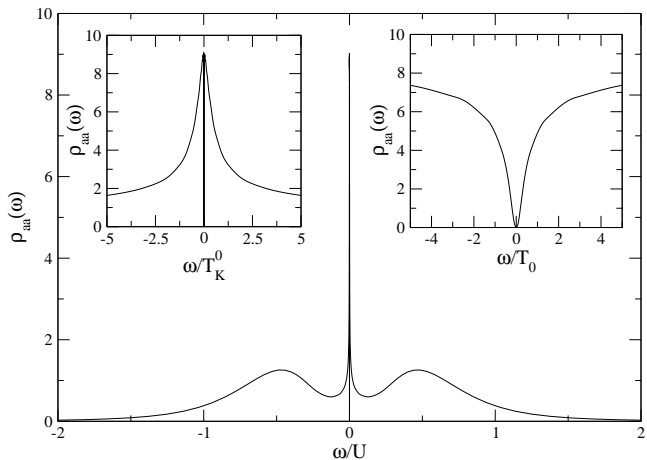}
\caption{Spectral density of dot $a$ at $T=0$. Parameters are
$U=0.5$, $\Delta=0.035$ and $t_{ab}=0.0022$. The broadened
atomic levels of dot $a$ are clearly seen at $\omega = \pm U/2$.
The left inset shows the central Kondo peak of width $T_K^0\sim
7\times 10^{-4}$. The right inset shows the Kondo hole of width $T_0\sim
1.78\times 10^{-10}$.} \label{fig:3scale}
\end{figure}

Figure~\ref{fig:3scale} shows the spectral density on dot $a$ at
zero temperature in the symmetric case for $U=0.5$ ,
$\Delta=0.035$ and $t_{ab}= 0.0022$. This value corresponds to the
third curve from the top in Fig.~\ref{fig:fig1}.

We observe three features,
a very narrow central peak and two broad Coulomb peaks located
at $\omega = \pm U/2$. We determine the Kondo scale $T^0_K =
7\times 10^{-4}$ from the full width at half-maximum of the central feature.
The insets to the figure are zooms of the central peak. The
 left inset shows the Kondo
peak plotted as a function of the reduced frequency
$\omega/T^0_K$.
The right inset shows the
 Kondo hole in the density of states present over the tiny
frequency interval $\left| \omega \right| < T_0 = 1.78\times 10^{-10}$.
 This feature is obviously not visible on the scales of either
 the main plot or the left inset.

\begin{figure}[tbp]
\includegraphics[width=8.5cm,clip=true]{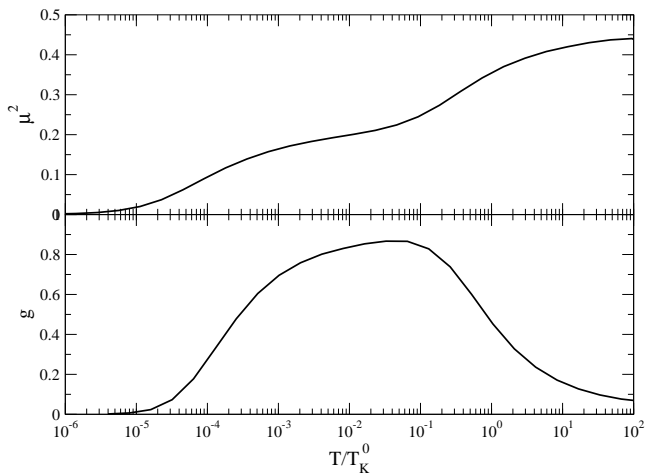}
\caption{Upper panel: Magnetic moment of the DQDD at zero field
as a function of
temperature for  $t_{ab}=0.003$. The other parameters are as in
Fig.\ref{fig:fig1}. Lower panel:
Conductance through the device at zero field as a function
of the temperature. Three regimes can be clearly seen both in the
conductance and the magnetic moment.} \label{fig:fig3}
\end{figure}

The conductance of the DQDD is entirely determined by the spectral
density of dot $a$ [cf. Eq. (\ref{eq:cond})]. The temperature
 dependence of $g$ in zero magnetic
field is represented in the lower panel of
Fig.~\ref{fig:fig3} for $t_{ab}=3\times 10^{-3}$ with the rest of
 the parameters fixed at the values given above.
For temperatures $T>T_K^0$ the device is in the Coulomb blockade
regime and the conductance is low.
 With decreasing temperature the Kondo correlations start
to build up and the magnetic moment of dot $a$ decreases as shown
in the upper panel of the figure.  For $T\lesssim T_K^0$ the spin
on dot $a$ is quenched resulting in resonant scattering through
dot $a$ and an enhanced conductance. For temperatures $T\lesssim
T_0 \sim 7\times 10^{-8}$, however, the spin of the second dot becomes
also screened and the Kondo hole  appears
 in the density of states. In this regime the
conductance decreases again.

\begin{figure} [tbp]
\includegraphics[width=8.5cm,clip=true]{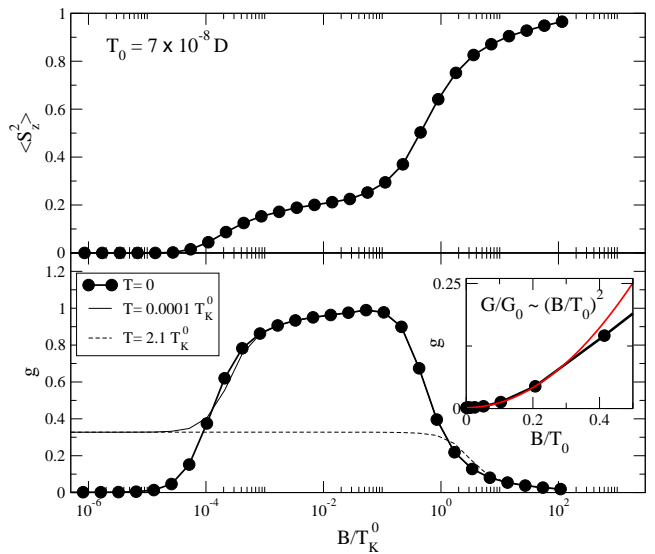}
\caption{Upper panel : magnetic moment of the DQDD as a function 
of magnetic field at $T=0$. Lower panel: conductance as a function of the
 magnetic field  at low temperatures. Inset: Field-dependence of
 the $T=0$ conductance. The coupling between the dots $t_{ab}=0.003$.
The other parameters are as in
Fig.\ref{fig:fig1}.} \label{fig:GdB}
\end{figure}

We now turn to a description of our results
in the presence of a magnetic field.
Data are displayed in Fig.~\ref{fig:GdB} for the same parameters
 as in Fig.~\ref{fig:fig3}.
The upper panel shows the field-dependence of  the
 square of $z$-component of the
 total spin ${\bf S = S_a + S_b}$ of the DQDD at $T=0$.
 The lower panel shows that of the conductance. 

For fields $B \gg T^0_K$ the spins of both dots are fully
 polarized and $\langle S_z^2\rangle  = 1$. The Kondo effect can 
 not take place and the conductance is small. 

For $B <  T^0_K$ the first stage of the
Kondo effect takes place and the spin on dot
 $a$ is quenched. At this point 
$\langle S_z^2\rangle  \sim  1/4 $ indicating that
 the spin on dot $b$ still remains polarized by the field.
 In this region of fields 
 the conductance increases with decreasing field, approaching
 its maximum value, $g=1$.

Upon further decrease of the field,
 Kondo screening of the second spin starts and, 
 for  $B\lesssim T_0 \approx 10^{-4}\times T^0_K$,
 it is in turn fully quenched. Then, the
 conductance decreases again and vanishes as $B\to 0$.

The inset to the figure shows that the low-field conductance varies
 quadratically with $B$ with a characteristic field $B^\star \sim T_0$
in full agreement with Eqs.~(\ref{g-de-B}) and (\ref{chiSBMFT}).

The effect of the temperature on the
 field-dependence of the conductance is also shown in the figure. For
 $T = 10^{-4}\;T^0_K \approx T_0$ and fields $B > T$
 the results are very similar to those obtained at zero temperature.
 At lower fields, however, the conductance saturates.
 This behavior arises because, at this temperature $T \gtrsim T_0$, 
 quenching of  spin $b$ is only partial and
 the Kondo hole in the density of
states ceases to develop. For $T > T^0_K$ both stages of the Kondo
effect are suppressed by thermal fluctuations and the conductance is
featureless as a function of $B$.


\section{Conductance as a function of the gate voltage} 
\label{sec:g-de-Vg-T}
In this Section we discuss the dependence of the conductance upon the gate voltage. For simplicity we restrict the analysis to the case of identical dots  $\varepsilon_a=\varepsilon_b=\varepsilon$, $U_a=U_b=U$. We have
checked, however, that the 
results for other situations are qualitatively similar.

\begin{figure}[tbp]
\includegraphics[width=8.5cm,clip=true]{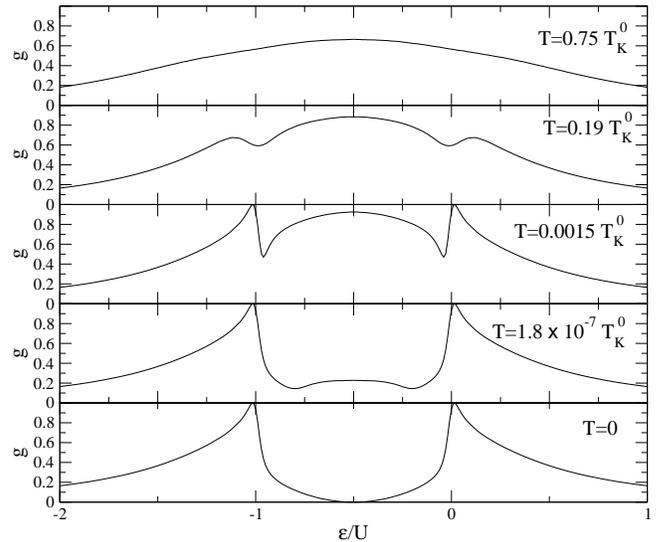}
\caption{Conductance through the double dot system as a function
of  the gate voltage $V_g=-\varepsilon_a=-\varepsilon_b=-\varepsilon$ and
different temperatures. Parameters are $U_a=U_b=0.25$,
 $t_{ab}=0.025$ and
$\Delta=0.125$.} \label{fig:Vgtemp}
\end{figure}

We display in Fig.~\ref{fig:Vgtemp} the conductance as function of
 $\varepsilon$, calculated from Eq.~(\ref{eq:cond}), for
 the set of parameters  $U=0.25,t_{ab}=0.025$ and $\Delta=0.125$. 
In this case the Kondo temperature and the effective magnetic coupling 
between the dots are  $T^0_K \sim \Delta$ and $J_{ab} \approx 0.01 < T^0_K$ for
 $\varepsilon = -U/2$.

At $T=0$ we observe two peaks of perfect conductance, $g=1$,
 and a wide region of low conductance between the two.
The conductance vanishes  at $\varepsilon = -U/2$ 
as discussed in the previous Section. 

We have checked that the conductance at $T=0$ 
satisfies Luttinger's theorem by 
computing $g$ from Eq.~(\ref{eq:condFL}) using
 the charge of the DQDD obtained from the NRG calculation
 and comparing it with the outcome of the direct calculation.
 The two results are the same within the numerical uncertainty.

The peaks of perfect conductance $g=1$ occur when the DQDD is in either of 
the charge states $n_a+n_b=1$ or $n_a+n_b=3$. At large $\varepsilon$
 the  charge of the DQDD is very small and increases with
 decreasing $\varepsilon$. It reaches $n=1$ near  
 $\varepsilon^\star$ such that  
the lowest energy level of the $a-b$ ``molecule'' crosses the Fermi level.
In the absence of interactions this is 
 $\varepsilon^\star = t_{ab} $. Interaction effects will reduce
 $\varepsilon^\star$, an effect that is already present for  
 weak interactions for which a simple Hartree-Fock aproximation
 leads to $\varepsilon^\star =
 \sqrt{t_{ab}^2 + \left(U \delta /4\right)^2} - U/4$,
 where $\delta = n_b - n_a$ is the difference between the charges on the two dots. For our paramenters the reduction is very large and we find 
 $\varepsilon^\star \approx 0.001 \ll t_{ab}$.

For temperatures  $0< T <T_K^0$, the conductance
increases at the center of the valley due to the disappearance of the
 second stage of the Kondo efect. This leads to the three peak
structure observed in the second and third panels from the bottom. 
If the temperature is further increased, the height of the conductance peaks 
decreases and, for temperatures $T\sim \Delta$, 
there is a single broad conductance peak. For our values of the parameters there are no Coulomb blockade peaks because 
their width $\Delta$ is comparable with their separation $U$.

\begin{figure}[tbp]
\includegraphics[width=8.5cm,clip=true]{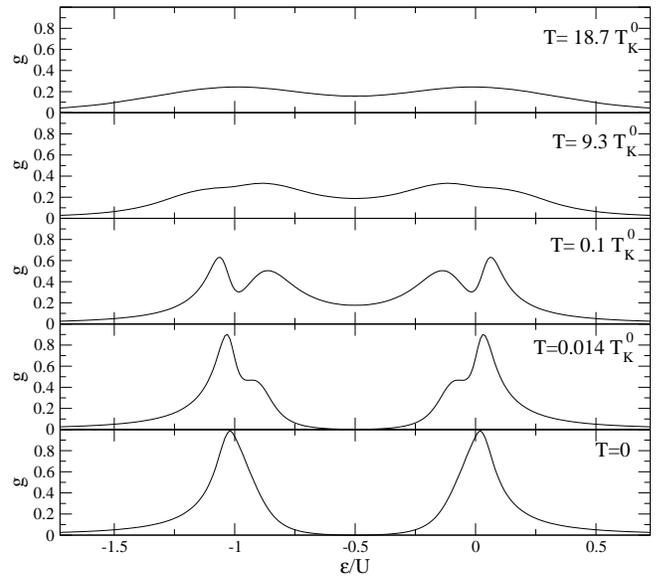}
\caption{Same as in Fig.~\ref{fig:Vgtemp} but for a different set
of parameters: $U_a=U_b=0.5$, $t_{ab}=0.05$ and $\Delta=0.063$.}
\label{fig:Vgtemp1}
\end{figure}

Figure~\ref{fig:Vgtemp1} shows the function
 $g(\varepsilon)$ at several temperatures for a different set of parameters,
 $U_a=U_b=0.5$, $t_{ab}=0.05$, and $\Delta=0.063$.
 In this case, $T^0_K \sim 0.01$ and $J_{ab} \sim 0.02 > T^0_K$. 
This a case in which the physics at the symmetric point is dominated 
by the formation of a strongly bound singlet between the spins of the 
two dots.  The behavior at zero temperature is qualitatively similar to
 that found in the previous case. The evolution of the conductance
 with temperature is however quite different on two accounts. First, 
 the conductance at $\varepsilon = - U/2$ is always a minimum; second,
 two additional conductance peaks appear upon rasing the temperature in the
 region $0 < T < T^0_K$. This phenomenon can be understood by noticing
 that while both $J_{ab}$ and $T^0_K$ increase with 
$\varepsilon$ in the valley away from $\varepsilon=-U/2$, the increase 
 of $T^0_K$
 is more pronounced because, in this case, the dependence on $\varepsilon$
 appears in the exponential. As result, while $J_{ab} > T^0_K$ for
 $\epsilon = 0$ there is an inversion of the inequality for sufficiently
 large $\epsilon$. When this inversion takes place the normal Kondo
 effect can occur and results in an enhancement of the conductance
 around some intermediate point. Finally, at high temperatures
 the usual Coulomb blockade peaks are this time visible because, for this parameter set, $\Delta \ll U$.         

\begin{figure}[tbp]
\includegraphics[width=8.5cm,clip=true]{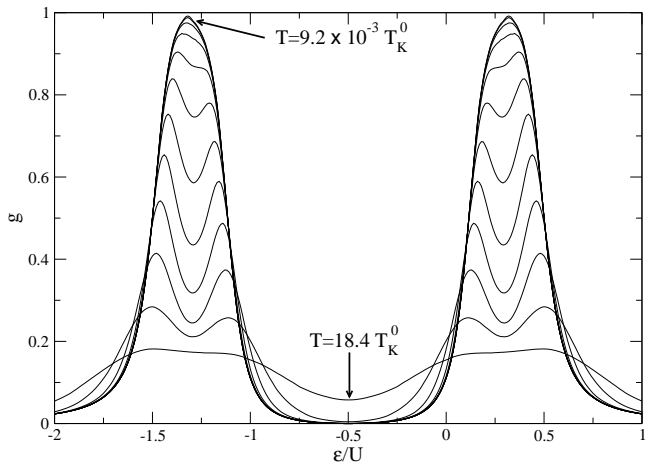}
\caption{Same as in Fig.~\ref{fig:Vgtemp1} but for $t_{ab}=0.25$.
 The curves correspond to temperatures ranging from
 $9.2\times 10^{-3} T^0_K$
 to  $18.4\;T^0_K$ from bottom to top (in the central region).
 Each curve corresponds to a temperature that is twice as high as that
 of the previous one.}
\label{fig:Vgtemp2}
\end{figure}

We close this Section with an analysis of a case in
 which the hopping matrix element between the dots is large. 
Fig.~\ref{fig:Vgtemp2} shows results for  
 $t_{ab} = 0.25=U/2$
 with all the other parameters as in Fig.~\ref{fig:Vgtemp1}.
 The appropriate starting 
 point for a qualitative description of this case 
 is an isolated $a-b$ molecule whose bonding and antibonding
 orbitals have energies $\varepsilon_{\pm}= \varepsilon \pm t_{ab}$.
 Consider the case in which the energy of the bonding orbital crosses
 the Fermi energy. The probability of occupation of the antibonding
 orbital is now strongly reduced 
because of the large energy gap $2 t_{ab}$ that separates its energy from that 
of the bonding state. States with non-zero occupation of
  the antibonding state can then safely be projected out of the
 subspace of available states using a generalized 
Schrieffer-Wolff transformation 
just as one projects out states with double occupancy of either of the dots.
 The resulting effective problem describes a {\it single} quantum dot with
 renormalized parameters. We thus expect to observe 
 the ordinary single-dot Kondo effect around
 $\varepsilon = t_{ab}$ and  $\varepsilon = - t_{ab} - U$. 
 This is precisely what  the figure shows. At low $T$, there are two 
peaks  of width $\sim U$ in the conductance 
in the range of  values of $\varepsilon$
for which the occupation number of the DDQD is odd.
 In the peak region we observe Coulomb blockade and
 the Kondo effect depending on the temperature.
~\footnote{We thank C. A. B\"usser for pointing out to us that this effect was likely to occur.}

\section{Conclusions} 
\label{sec:conc}
In summary, we studied the spectral, magnetic and transport properties
 of a double quantum-dot device in which one of the dots is coupled
 to perfect conducting leads.

The zero-temperature conductance of the system vanishes
 when the total charge of the device is an even integer and it is 
perfect when it is an odd integer.

 When both dots are individually in the Kondo regime 
two different physical situations are possible depending on whether
 the magnetic coupling between the dots is smaller or larger than 
the Kondo temperature of a single dot. 
When $J_{ab} \ll T^0_K$ the transport properties of the system 
reflect 
 the existence of  
 a two-stage Kondo effect. In the first one, the magnetic moment of the 
dot connected to the leads is screened at $T^0_K$, the Kondo temperature
 of the isolated dot. In this regime the conductance increases with
 decreasing temperature. At temperature $T_0 \ll T^0_K$ the 
magnetic moment of the second dot is also quenched, a process that
 leads to a conductance decreasing with temperature. 
 A similar non-monotonic behavior is observed in  
the field-dependence of the conductance. 
In the opposite case, $J_{ab} \gg T^0_K$, the magnetic moments of the dots
 bind forming a strong  singlet weakly coupled to the leads and the 
 low temperature conductance decreases monotonically with $T$.

Application of a gate voltage leads to a rich variety of behaviors 
 for the dependence of  the conductance upon gate voltage and temperature. 
A general feature is that the conductance is small in a very wide range 
of values of $V_g$ at all temperatures. At low temperatures, regions of
 nearly perfect conductance do exist, but they are restricted to narrow
 intervals of values of the gate voltage.
Phenomena similar to some of those reported here have been found in a study
 of a multilevel quantum dot. \cite{Hofstetter2002}

In a recent paper \cite{Apel2004} (see also Ref.
\onlinecite{Busser2004b}) the same model was studied using exact
diagonalization methods in small clusters. These authors investigate the
 zero-temperature 
conductance of the device as a function of
 the positions of the dot's energy levels $\varepsilon_{a,b}$.
 In the regime of small
$t_{ab}$ they report perfect conductance for
$\varepsilon_a=\varepsilon_b=-U/2$. 
These numerical results are inconsistent with ours and, most importantly, 
with the implications of Luttinger's theorem. 
The exact diagonalization method apparently  
 fails to capture the second stage of the 
Kondo effect responsible for the 
supression of the conductance at zero temperature. 

The reason for this failure is easy to understand.  
In order to correctly describe the single dot Kondo effect
 in a finite cluster, the level
spacing in the leads must be smaller than the Kondo temperature.
\cite{Thimm1999,Affleck2001,Cornaglia2002a,Simon2002,Cornaglia2003b}
 For a QD coupled to a linear chain the number of sites required is $N_s
\gg 2D/T_K$ which becames numerically intractable beyond $N_s \sim 20$.
Large Kondo temperatures must then be considered. The Kondo temperature of the 
second dot is much lower than $T_K$ rending
the problem intractable by the exact diagonalization method
 in the limit of small $t_{ab}$.
However, these zero-temperature exact diagonalization results describe qualitatively the behavior for temperatures $T_0< T \ll T^0_K$ or for large $t_{ab}$.

Finally, we want to stress that our conclusions are valid in the case
 in which the effective magnetic coupling between the dots $J_{ab}$ is
 antiferromagnetic and do not apply directly to the case of
 ferromagnetic coupling. 

\section{Acknowledgments}
We thank C. A. B\"usser and E. Dagotto for useful correspondence.


\end{document}